\documentclass[twoside,twocolumn,english,prb,twocolumn,superscriptaddress,showpacs,amsmath]{revtex4}
\usepackage[T1]{fontenc}
\usepackage[latin9]{inputenc}
\usepackage{amsmath}
\usepackage{graphicx}
\usepackage{amssymb}
\usepackage{esint}

\makeatletter

\providecommand{\tabularnewline}{\\}

\@ifundefined{textcolor}{}
{%
 \definecolor{BLACK}{gray}{0}
 \definecolor{WHITE}{gray}{1}
 \definecolor{RED}{rgb}{1,0,0}
 \definecolor{GREEN}{rgb}{0,1,0}
 \definecolor{BLUE}{rgb}{0,0,1}
 \definecolor{CYAN}{cmyk}{1,0,0,0}
 \definecolor{MAGENTA}{cmyk}{0,1,0,0}
 \definecolor{YELLOW}{cmyk}{0,0,1,0}
 }
\newenvironment{lyxlist}[1]
{\begin{list}{}
{\settowidth{\labelwidth}{#1}
 \setlength{\leftmargin}{\labelwidth}
 \addtolength{\leftmargin}{\labelsep}
 }}
{\end{list}}

\tolerance7000\newcommand{\be}{\begin {equation}}\newcommand{\ee}{\end {equation}}\newcommand{\bear}{\begin{eqnarray}}\newcommand{\eear}{\end{eqnarray}}\topmargin0.1cm

\makeatother

\usepackage{babel}

\makeatother

\usepackage{babel}

\makeatother

\usepackage{babel}

\begin{document}

\title{Theory of Electro-Optical Properties of Graphene Nanoribbons}

\author{Kondayya Gundra}

\altaffiliation{Permanent Address: Theoretical Physics Division, Bhabha Atomic Research Centre, Mumbai 400085, INDIA}

\email{naiduk@barc.gov.in, shukla@iitb.ac.in}

\author{Alok Shukla}

\affiliation{Department of Physics, Indian Institute of Technology, Bombay, Mumbai
400076 INDIA}
\begin{abstract}
We present calculations of the optical absorption and electro-absorption
spectra of graphene nanoribbons (GNRs) using a $\pi-$electron approach,
incorporating long-range Coulomb interactions within the Pariser-Parr-Pople
(PPP) model Hamiltonian. The approach is carefully bench marked by
computing quantities such as the band structure, electric-field driven
half metallicity, and linear optical absorption spectra of GNRs of
various types, and the results are in good agreement with those obtained
using\emph{ ab initio }calculations. Our predictions on the linear
absorption spectra for the transversely polarized photons provide
a means to characterize GNRs by optical probes. We also compute the
electro-absorption spectra of the zigzag GNRs, and argue that it can
be used to determine, whether or not, they have a magnetic ground
state, thereby allowing the edge magnetism to be probed through non-magnetic
experiments. 
\end{abstract}


\maketitle

\section{Introduction}

\label{sec:intro}

Discovery of graphene \cite{Novoselov-1} has stimulated intense research
in the field from the point-of-view of both fundamental physics, and
promising applications \cite{Novoselov-2,Navoselov-3,Zhang}. Of particular
interest are recently synthesized\cite{Han} quasi-one-dimensional
(1D) nanostructures of graphene called graphene nanoribbons (GNRs)
which have technologically promising electronic and optical properties
because of the confinement of electrons owing to the reduced dimensions.
As a result, numerous theoretical studies of electronic, transport,
and optical properties of GNRs of various type have been performed
over the years\cite{fujita,Nakada,okada,Ezawa,SonYW,Son,Scuseria,yang2,reichl-prb,yang,Prezzi}.
The structural anisotropy of GNRs must exhibit itself in an anisotropic
optical response with respect to the photons polarized along the length
of the ribbons ($x$ polarized, or longitudinally polarized) as against
those polarized perpendicular to it ($y$ polarized or transversely
polarized), with GNRs being in the $xy$-plane. Despite its obvious
importance, anisotropy in the optical response of GNRs has not been
studied in any of the reported optical absorption calculations, which
concentrate only on the longitudinal component of the spectra\cite{Scuseria,yang2,reichl-prb,yang,Prezzi}.
In this work we study this anisotropy in detail, and make predictions
which can be tested in optical experiments on oriented samples of
GNRs, and can serve as a means for their optical characterization.

Electro-absorption (EA) spectroscopy, which consists of measuring
optical absorption in the presence of a static external electric (E)
field, has been used extensively to probe the electronic structure
and optical properties of conjugated polymers and other materials\cite{EA-spec}.
GNRs, being $\pi$-conjugated systems, will also be amenable to similar
EA probes, and, therefore, we have calculated the EA spectrum of zigzag
GNRs (ZGNRs) in this work. ZGNRs have been predicted to possess a
magnetic ground state, with oppositely oriented spins localized on
the opposite zigzag edges of the ribbons\cite{fujita,okada}. Our
calculated EA spectra of ZGNRs depends strongly on whether, or not,
they exhibit edge magnetism, thereby, allowing its detection by optical
means.

Most of the theoretical approaches used to study the electronic structure
of GNRs are broadly based upon: (a) tight-binding method,\cite{Nakada,fujita,Ezawa}
(b) Dirac equation approach, derived using the linearity of the band
structure in the region of interest,\cite{graphene-rmp} (c) \emph{ab
initio} DFT and GW based approaches,\cite{Son,SonYW,yang2,yang} and
(d) Hubbard model based approaches.\cite{jung-hubbard,yazyev-hubbard,fernandez-hubbard,ext-hub}
But, it is obvious from the chemical-structure of graphene and GNRs
that the electrons close to the chemical potential are itinerant $\pi$-electrons
which determine their low-energy excitations. In $\pi$-electron systems
such as various aromatic molecules and conjugated polymers, it is
well-known that role of electron-electron (e-e) interactions cannot
be ignored when describing their electronic properties.\cite{salem-book}Therefore,
it is inconceivable that the long-range e-e interactions will be insignificant
in graphene and related structures. The effective $\pi$-electron
approaches such as the Pariser-Parr-Pople (PPP) model Hamiltonian,\cite{ppp}
which incorporate long-range e-e interaction, have been used with
considerable success in describing the physics of $\pi$-conjugated
molecules and polymers.\cite{salem-book} Computationally speaking,
PPP model has the advantage of including the long-range Coulomb interactions
of $\pi$ electrons within a minimal basis, thereby allowing calculations
on such systems with limited computer resources, as compared to the
ab initio approaches. Indeed, in our earlier works we have used the
PPP model to extensively to study the electronic structure and optical
properties of \emph{finite} $\pi$-electron systems such as conjugated
molecules and oligomers at various levels of theory.\cite{shukla}
Therefore, in this work, we have decided to extend our PPP model based
approach to study the physics of GNRs in the bulk limit. Because,
to the best of our knowledge, this is the first application of the
PPP model to the GNR physics, we have carefully bench marked it for
quantities such as the band structure, electric-field driven half
metallicity, and linear optical absorption spectra against the published
\emph{ab initio} works on GNRs, and the results are in very good agreement
with each other. 

The remainder of this paper is organized as follows. In the next section,
we outline the theoretical aspects of our work. In section \ref{sec:results},
we present and discuss our results. Finally, in section \ref{sec:summary}
we present our conclusions and discuss the directions for the future
work.

\section{Theoretical Details}

\label{sec:theory}

The PPP model Hamiltonian,\cite{ppp} with one $\pi$-electron per
carbon atom (half-filled case), is given by\begin{eqnarray}
H=-\sum_{i,j,\sigma}t_{ij}(c_{i\sigma}^{\dagger}c_{j\sigma}+c_{j\sigma}^{\dagger}c_{i\sigma})+\nonumber \\
U\sum_{i}n_{i\uparrow}n_{i\downarrow}+\sum_{i<j}V_{ij}(n_{i}-1)(n_{j}-1)\label{eq:ham-ppp}\end{eqnarray}
 above $c_{i\sigma}^{\dagger}$ creates an electron of spin $\sigma$
on the $p_{z}$ orbital of carbon atom $i$, $n_{i\sigma}=c_{i\sigma}^{\dagger}c_{i\sigma}$
is the number of electrons with spin $\sigma$, and $n_{i}=\sum_{\sigma}n_{i\sigma}$
is the total number of electrons on atom $i$. The parameters $U$
and $V_{ij}$ are the on--site and long--range Coulomb interactions,
respectively, while $t_{ij}$ is the one-electron hopping matrix element
which, if needed, can be restricted to nearest-neighbors(NN). On setting
$V_{ij}=0$, the Hamiltonian reduces to the Hubbard model. The parametrization
of the Coulomb interactions is Ohno like \cite{Ohno},\begin{equation}
V_{i,j}=U/\kappa_{i,j}(1+0.6117R_{i,j}^{2})^{1/2}\;\mbox{,}\label{eq-ohno}\end{equation}

where, $\kappa_{i,j}$ depicts the dielectric constant of the system
which can simulate the effects of screening, and $R_{i,j}$ is the
distance in \AA{}~ between the $i$-th and the $j$-th carbon atoms.
The Hartree-Fock (HF) theory for periodic one-dimensional systems,
within the linear combination of atomic orbitals (LCAO) approach is
fairly standard, and we have implemented both its restricted (RHF)
and unrestricted (UHF) variants. The lattice sums are performed in
the real space by including a large number of unit cells, and integration
along the Brillouin Zone (BZ) was performed using the Gauss-Legendre
quadrature approach \cite{hf-pol}. The convergence with respect to
the numbers of unit cells included in the lattice sums, as well as
$k$-points used for BZ integration, was carefully checked. 

Our calculations, to the best of our knowledge, are the first applications
of the PPP model to GNRs in the \emph{bulk} limit; therefore, it is
important to obtain a suitable set of Coulomb parameters for these
systems. In our previous calculations on conjugated molecules and
polymers \cite{shukla}, we used two sets of Coulomb parameters namely
(a) {}``standard parameters'' with $U=11.13$ eV and $\kappa_{i,j}=1.0$,
and (b) {}``screened parameters'' with $U=8.0$ eV and $\kappa_{i,j}=2.0$
($i\neq j)$ and $\kappa_{i,i}=1$, proposed initially by Chandross
and Mazumdar to study phenyl-based conjugated polymers.\cite{chandross}
In the absence of extensive experimental data, we adopted the criterion
of good agreement between the \emph{ab initio} GW band gaps of armchair
GNRs (AGNRs)\cite{yang2} and our PPP band gaps, to choose the Coulomb
parameters. The tuning of the parameters was done for AGNR-12 (AGNR-$N_{A}$,
denoting an AGNR with $N_{A}$ dimer lines across the width), and
with a modified set of screened parameters ($U=6.0$ eV,$\kappa_{i,j}=2.0$
($i\neq j)$ and $\kappa_{ii}=1$), and NN hopping matrix element
$t_{1}=-2.7$ eV. As a result, good agreement was obtained for AGNR-12
between the PPP band gap (1.75 eV) and the corresponding GW value
of Yang \emph{et al}.\cite{yang2}. Therefore, we have decided to
use these modified Coulomb parameters throughout these calculations,
with the aim that they will incorporate the GW-level electron-correlation
results implicitly in our results.

\section{Results and Discussion}

\label{sec:results}

\begin{figure}
\begin{lyxlist}{00.00.0000}
\item [{\includegraphics[width=6cm]{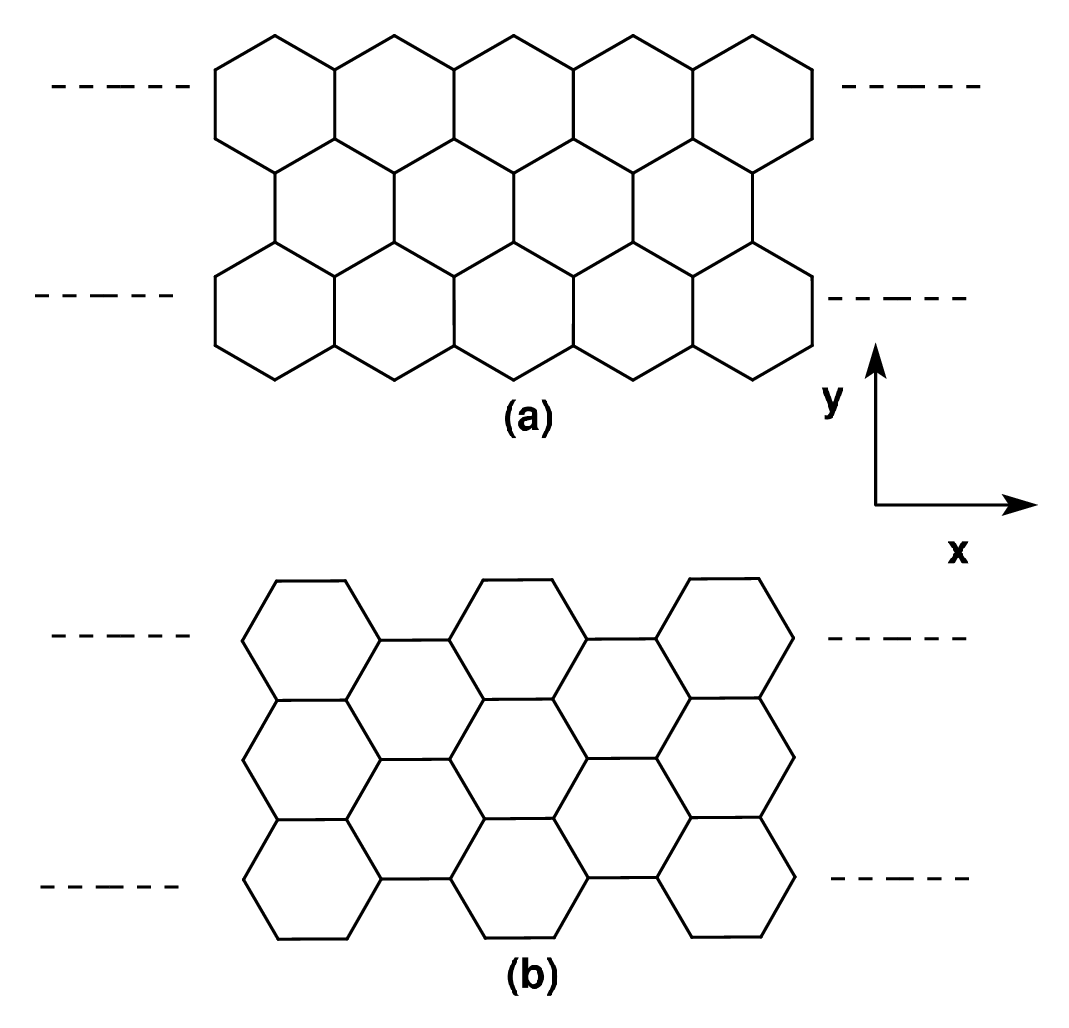}}]~
\end{lyxlist}
\caption{The structures of (a) a ZGNR and (b) an AGNR. The ribbons are assumed
to lie in the $xy$ plane, with the periodicity in the $x$ direction. }

\label{Fig:GNRs}
\end{figure}

The schematic structures of AGNRs and ZGNRs studied in this work are
presented in Fig. \ref{Fig:GNRs}. Next, we present the results of
our PPP model based calculations on various quantities, for AGNRs
and ZGNRs.

\subsection{Band Structure}

\begin{figure}
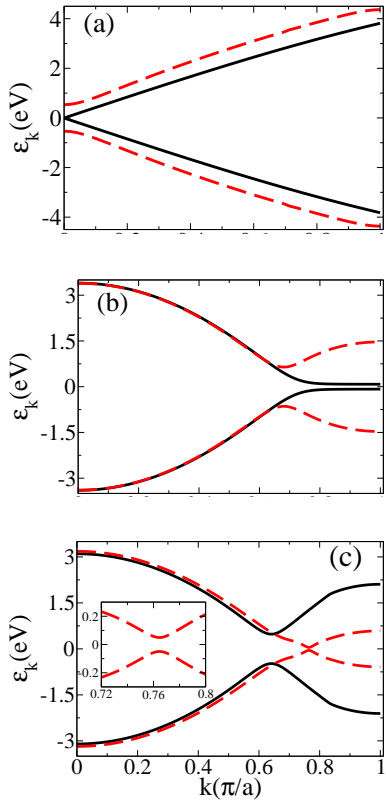

\includegraphics[width=5cm]{fig2a}

\includegraphics[width=5cm]{fig2b}

\includegraphics[width=5cm]{fig2c}

\caption{(Color online) Band structure near the Fermi energy ($E_{F}=0$) of:
(a) AGNR-11 obtained using the Hubbard Model (black solid line), with
$U=2.0$, and the PPP-RHF approach (red broken line), (b) ZGNR-12
obtained using the PPP-RHF approach for the non magnetic state (black
solid line), and the PPP-UHF approach for the magnetic state (red
broken line) in which the bands of up and down spins are degenerate,
(c) ZGNR-16, obtained using PPP-UHF model, in the presence of a lateral
electric field of 0.1 V/\AA\  so that the up-down degeneracy is
lifted (red broken/black solid bands represents up/down spins) with
$E_{g}^{m(up)}=0.11$ eV (magnified in the inset), and $E_{g}^{m(down)}=0.97$
eV.}

\label{Fig:band-GNR}
\end{figure}

In Fig. \ref{Fig:band-GNR} (a) we present the band structures of
AGNR-11 obtained using the Hubbard model with $U=2.0$, and the PPP
model. At the tight-binding level all the AGNRs with $N_{A}=3p+2$
($p$ a positive integer) are predicted to be gapless. However, \emph{ab
initio} DFT calculations predict all types of AGNRs to be gapped,
including those with $N_{A}=3p+2$\cite{Son,Scuseria}. Our RHF calculations
are in agreement with the DFT results, and also predict all families
of AGNRs, including $N_{A}=3p+2$ to be gapped, as is obvious from
our PPP results for AGNR-11 presented in Fig. \ref{Fig:band-GNR}
(a). The noteworthy point is that the Hubbard model, with the currently
accepted values of $U$ predicts a negligible gap for $N_{A}=11$
(\emph{cf}. Fig. \ref{Fig:band-GNR}), a result in complete disagreement
with the DFT, and our PPP results. Thus, from this case it is obvious
that for AGNRs, long-range Coulomb interactions as included in the
PPP model play a very important role of opening up the gap for the
$N_{A}=3p+2$ case. Our PPP value of the band gap 1.06 eV of this
AGNR is again in excellent agreement with the \emph{ab initio} GW
result reported by Yang \emph{et al}.\cite{yang2}.

The case of the ground state of ZGNRs is an interesting one with several
authors reporting the existence of a magnetic ground state, with oppositely
oriented spins localized on the opposite zigzag edges of the ribbons\cite{fujita,okada},
a result verified also in several first principles DFT calculations\cite{Son,yang}.
We investigated this in our PPP model calculations by using the RHF
method for the nonmagnetic state and the UHF method for the magnetic
one, and the results are summarized in Table \ref{tab:energy-zgnr}.
We find that for a ZGNR of width $N_{Z}$ ($N_{Z}$ $\equiv$ number
of zigzag lines across the width), ZGNR-$N_{Z}$ in short, the total
energy/cell of the magnetic state ($E_{m}$) is lower as compared
to that of the non-magnetic ($E_{nm}$) one, and the energy difference/atom
between the non-magnetic and the magnetic states ($\Delta E=(E_{m}-E_{nm})/N_{at},$
$N_{at}\equiv$number of atoms in the unit cell) decreases with the
increasing ribbon-width, consistent with the non-magnetic ground state
of graphene. The band gap for the magnetic state ($E_{g}^{m}$) is
much larger than that of the non-magnetic one ($E_{g}^{nm}$). The
non-zero gaps obtained for the non-magnetic states of ZGNRs is an
artifact of the RHF approach. The band structures of the magnetic
and non-magnetic states of ZGNR-12 computed using the PPP model are
presented in Fig. \ref{Fig:band-GNR} (b) , and it is obvious that,
for the magnetic case, our results are qualitatively very similar
to the reported \emph{ab initio} band structures\cite{Son,yang}.
Quantitatively speaking, for ZGNR-8, we obtain $E_{g}^{m}=1.70$ eV,
which is higher than the reported GW value of 1.10 eV \cite{yang}.
Our band gap for AGNR-11 was in excellent agreement with the GW value,
but that is not the case with ZGNRs. We believe that it could possibly
be because: (a) our Coulomb parametrization was based upon \emph{ab
initio} GW results\cite{yang2} on an AGNR, and (b) electron-correlation
effects are stronger in ZGNRs as compared to AGNRs, and the HF approach
adopted here ignores those effects. 

\begin{table}
\begin{tabular}{|c|c|c|c|c|c|}
\hline 
Width & \multicolumn{3}{c|}{Total Energy (eV)} & \multicolumn{2}{c|}{Band gap (eV)}\tabularnewline
\hline
\hline 
$N_{Z}$ & $E_{nm}$ & $E_{m}$ & $\Delta E$ & $E_{g}^{nm}$ & $E_{g}^{m}$\tabularnewline
\hline 
4  & -23.059  & -23.261 & -0.025 & 0.524 & 2.414 \tabularnewline
\hline 
6  & -35.559 & -35.825 & -0.022 & 0.336  & 2.005\tabularnewline
\hline 
8 & -48.103 & -48.403 & -0.019 & 0.246 & 1.694 \tabularnewline
\hline 
12 & -73.237 & -73.570 & -0.014 & 0.161 & 1.287\tabularnewline
\hline 
16 & -98.417 & -98.743  & -0.010 & 0.046 & 1.037\tabularnewline
\hline
\end{tabular}

\caption{Variation of total energy/cell and the band gaps of ZGNR with the
width of the ribbon, computed using the PPP model. }

\label{tab:energy-zgnr}
\end{table}

In a pioneering work Son \emph{et al.}\cite{SonYW}, based upon \emph{ab
initio} DFT calculations, predicted that in the presence of a lateral
electric field, ZGNRs exhibit half-metallic behavior leading to their
possible use in spintronics. They demonstrated that for the field
strength 0.1 V/\AA, the gap for one of the spins of ZGNR-16 will
close, leading to metallic behavior for that spin orientation. In
Fig. \ref{Fig:band-GNR}(c) we present the band structure of the same
ZGNR exposed to the identical field strength, calculated using the
PPP model, and the tendency towards half-metallicity is obvious. While
 the band gap in the absence of the field was 1.037 eV, in the presence
of the field up-spin band gap is reduced to 0.11 eV, while the down-spin
gap decreases to 0.97 eV. Therefore, considering the fact that our
PPP model based approach does not incorporate electron-correlation
effects, its quantitative predictions are in very good agreement with
the \emph{ab initio} ones\cite{SonYW}, and thus it is able to capture
the essential physics of the electric-field driven half-metallicity
in ZGNRs.

\subsection{Optical Absorption}

Next we present the linear optical absorption spectra of GNRs, computed
within the PPP model. The optical absorption spectrum for the $x$-polarized
($y$-polarized) photons is computed in the form of the corresponding
components of the imaginary part of the dielectric constant tensor,
i.e., $\epsilon_{xx}^{(2)}$($\epsilon_{yy}^{(2)}(\omega)$), using
the standard formula\begin{equation}
\epsilon_{ii}^{(2)}(\omega)=C\sum_{v,c}\int_{-\pi/a}^{\pi/a}\frac{|\langle c(k)|p_{i}|v(k)\rangle|^{2}}{\{(E_{cv}(k)-\hbar\omega)^{2}+\gamma^{2}\}E_{cv}^{2}(k)}dk,\label{eq:eps2}\end{equation}
where $a$ is the lattice constant, $p_{i}$ denotes the momentum
operator in the $i$-th Cartesian direction, $\omega$ represents
the angular frequency of the incident radiation, $E_{cv}(k)=\epsilon_{c}(k)-\epsilon_{v}(k)$,
with $\epsilon_{c}(k)\:(\epsilon_{v}(k))$ being the conduction band
(valence band) eigenvalues of the Fock matrix, $\gamma$ is the line
width, while $C$ includes rest of the constants. We have set $C=1$
in all the cases to obtain the absorption spectra in arbitrary units.
The components of the momentum matrix elements $\langle c({\bf k})|{\bf {\bf p}}|v({\bf k})\rangle$
needed to compute $\epsilon_{ii}^{(2)}(\omega)$, for a general three-dimensional
system, can be calculated using the formula,\cite{opt-mat-el-2} 

\begin{eqnarray}
\langle c({\bf k})|{\bf {\bf p}}|v({\bf k})\rangle & = & \frac{m_{0}}{\hbar}\langle c({\bf k})|\nabla_{{\bf k}}H({\bf k})|v({\bf k})\rangle\nonumber \\
 &  & +\frac{im_{0}(\epsilon_{c}({\bf k})-\epsilon_{v}({\bf k}))}{\hbar}\langle c({\bf k})|{\bf d}|v({\bf k})\rangle,\label{eq:p-matel}\end{eqnarray}
where $m_{0}$ is the free-electron mass, $\nabla_{{\bf k}}H({\bf k})$
represents the gradient of the Hamiltonian (Fock matrix, in the present
case) in the ${\bf k}$ space, $\langle c({\bf k})|{\bf d}|v({\bf k})\rangle$
denotes the matrix elements of the position operator ${\bf d}$ defined
with respect to the reference unit cell, and accounts for the so-called
intra-atomic contribution.\cite{opt-mat-el-2}Note that Eq. \ref{eq:p-matel}
can also be used to compute the matrix element $\langle c({\bf k})|p_{y}|v({\bf k})\rangle$
needed to compute the absorption spectrum for the $y$-polarized light
for GNRs (which are periodic only in the $x$ direction), by setting
the first term on its right hand side to zero, because for a one-dimensional
system periodic along the $x$ direction, the Hamiltonian has no $k_{y}$
dependence. $\langle c({\bf k})|\nabla_{{\bf k}}H({\bf k})|v({\bf k})\rangle$
for the case of GNRs is obtained easily by calculating the numerical
derivative of the Fock matrix at various $k$-points of the one-dimensional
Brillouin zone. For the ${\bf d}$ operator, the usual diagonal representation
was employed. The calculation of the absorption spectra of the GNRs
for the $y-$polarized photons ($\epsilon_{yy}^{(2)}(\omega)$), to
the best of our knowledge, has not been done earlier. Because such
transverse excitations do not couple to the photons polarized along
the $x$ direction, they have also been called {}``dark excitons''
in the literature\cite{yang,yang2}. 

\begin{figure}
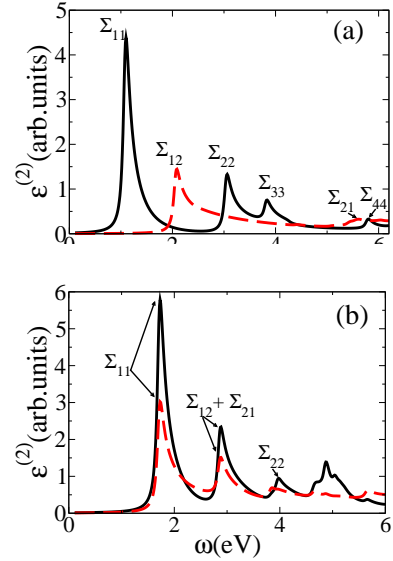

\includegraphics[width=5cm]{fig3a}

\includegraphics[width=5cm]{fig3b}

\caption{(Color online) Imaginary parts of the dielectric constant ($\epsilon_{xx}^{(2)}(\omega)$
in black solid, and $\epsilon_{yy}^{(2)}(\omega)$ in red broken lines)
computed using the PPP model, and modified screened parameters, for:
(a) AGNR-11, (b) ZGNR-8, with a magnetic ground state. Labels of the
peaks denote the bands involved in the transition (see text for an
explanation), and a line width of 0.05 eV was assumed throughout.}

\label{Fig:opt-a11}
\end{figure}

The optical absorption in AGNRs has been studied extensively by \emph{ab
initio} approaches in recent works \cite{yang2,Scuseria,Prezzi}.
In Fig. \ref{Fig:opt-a11}(a) we present the optical absorption spectrum
of the AGNR-11. If $\Sigma^{mn}$ denotes a peak in the spectrum due
to a transition from $m$-th valence band (counted from top) to the
$n$-th conduction band (counted from bottom), the peak of $\epsilon_{xx}^{(2)}(\omega)$
at 1.1eV is $\Sigma^{11}$, at 3.1 eV is $\Sigma^{22}$, at 3.8 eV
is $\Sigma^{33}$, and at 5.8 eV is $\Sigma^{44}$. The peaks of $\epsilon_{yy}^{(2)}(\omega)$
at 2.1 eV and 5.6 eV both correspond to $\Sigma^{12}$ and $\Sigma^{21}$.
The remarkable feature of the presented spectrum is that owing to
the symmetry of the AGNRs, the peaks corresponding to $x$- and $y$-polarized
photons are well separated in energy, and their relative intensities
can be measured by performing experiments on oriented samples. On
comparing our PPP spectrum ($\epsilon_{xx}^{(2)}(\omega)$) with the
\emph{ab initio} GW spectrum of Yang \emph{et al.} \cite{yang2},
we note that the locations of the first peaks close to 1.1 eV are
in excellent agreement with each other. However, our calculations
predict several higher energy peaks with significant intensities located
around 3.0 eV absent in the GW work. Furthermore, we also predict
the intensities of the $y$-polarized peaks, which was absent in the
work of Yang \emph{et al.} \cite{yang2}.

In Fig. \ref{Fig:opt-a11}(b) we present our calculated optical absorption
spectrum ($\epsilon_{xx}^{(2)}(\omega)$ and $\epsilon_{yy}^{(2)}(\omega)$)
for the ZGNR-8. The peaks in $\epsilon_{xx}^{(2)}(\omega)$ are located
at 1.7 eV ($\Sigma^{11}$), 2.9 eV ($\Sigma^{12}+\Sigma^{21}$), at
4.0 eV ($\Sigma^{22}$), while the prominent peaks of $\epsilon_{yy}^{(2)}(\omega)$
are at 1.7 eV ($\Sigma^{11}$) and 2.9 eV ($\Sigma^{12}+\Sigma^{21}$).
The noteworthy point is that most of the prominent peaks have mixed
polarization characteristics, unlike the case of AGNRs. This is because
of the fact that for magnetic ground states, the reflection about
the $xz$-plane is broken, leading to mixed polarizations. This is
an important result which can also be tested in oriented samples of
ZGNRs. Our PPP optical absorption spectrum of this ZGNR compares qualitatively
well to the GW spectrum computed by Yang \emph{et al}.\cite{yang},
although our peaks are consistently blue-shifted compared to the GW
result, due to the corresponding disagreement in the band structure.
Moreover, Yang \emph{et al}.\cite{yang} did not compute the peak
intensities for the $y$-polarized photons.

\subsection{Electro-Absorption}

\begin{figure}
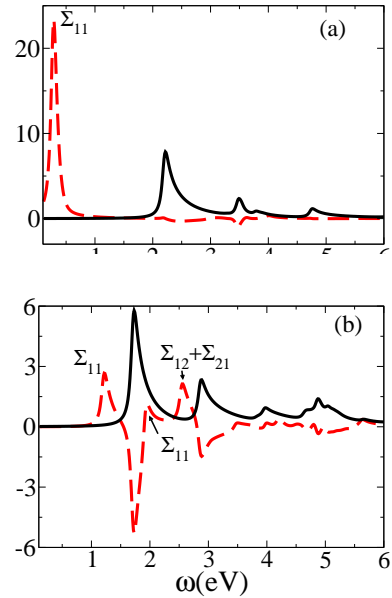

\includegraphics[width=5cm]{fig4a}

\includegraphics[width=5cm]{fig4b}

\caption{(Color online) Linear absorption spectrum (black solid) and electro
absorption (red broken) of ZGNR-8 for photons polarized along the
$x$ axis for: (a) non magnetic ground state, and (b) magnetic ground
state. A line width of 0.05 eV was assumed throughout, and the bands
involved in the electro-absorption peaks are indicated.}

\label{Fig:ele-rhf-z8}
\end{figure}

In Figs. \ref{Fig:ele-rhf-z8} we present the EA spectrum of ZGNR-8
computed as the difference of the linear absorption spectra with and
without an external static E-field of strength 0.1 V/\AA\ along
the $y$-axis. In Fig. \ref{Fig:ele-rhf-z8}(a) we present the EA
spectrum for the non-magnetic ground of ZGNR-8, computed using the
PPP-RHF approach. Without the external E-field, the $\Sigma_{11}$
transition is disallowed for the non-magnetic state of such a ZGNR
for the $x$-polarized light due to symmetry selection rules\cite{reichl-prb}.
However, in the presence of the field, due to the broken symmetry,
this transition becomes strongly allowed leading to a very strong
peak in the EA spectrum. Fig. \ref{Fig:ele-rhf-z8}(b) portrays the
EA spectrum of the same ZGNR for the magnetic ground state, and, here
the physics of half-metallicity manifests itself in that one observes
two energetically split peaks corresponding to two different $\Sigma_{11}$transitions
among up- and down-spin electrons. Thus, our calculations predict
that the EA signal is different for the ZGNRs depending on whether
they have a magnetic or a non-magnetic ground state, a result which
can be used to determine the nature of the ground state of ZGNRs using
EA spectroscopy. 

\section{Summary and Outlook}

\label{sec:summary}

In summary, we have used a PPP model based $\pi$-electron approach,
incorporating long-range Coulomb interactions, to study the electronic
structure and optical properties of GNRs in the \emph{bulk} limit.
In particular, we computed the optical absorption spectra of GNRs
for transversely polarized photons, in addition to the longitudinal
ones, thereby allowing us to investigate the anisotropic optical response
of these materials. Our predictions that for AGNRs longitudinal and
transverse polarized components will be well separated energetically,
while ZGNRs will exhibit absorption with mixed polarization, can be
tested in experiments on oriented samples. Furthermore, we also presented
first calculations of the EA spectra of ZGNRs, and our results suggest
a possibility of an optical determination of whether, or not, they
possess a ground state with edge magnetism. 

It will also be of interest to perform similar studies on bilayer
and other multilayer GNRs, to investigate how various properties of
the ribbons evolve, as the number of layers are increased. Of particular
interest is the case of multilayer ZGNRs, to probe as to what is the
nature of edge magnetism in those systems. Furthermore, it will also
be of interest to include excitonic effects in the optical absorption
spectrum of ZGNRs so as to perform a complete comparison with the
future experimental work on these systems. For the purpose, it is
important to go beyond the Hartree-Fock approach and include electron-correlation
effects. Work along all these directions is in progress in our group,
and the results will be communicated in the future publications.
\begin{acknowledgments}

\end{acknowledgments}
We thank the Department of Science and Technology (DST), Government
of India, for providing financial support for this work under Grant
No. SR/S2/CMP-13/2006. K. G is grateful to Dr. S. V. G. Menon (BARC)
for his continued support of this work.

\end{document}